\documentclass[a4paper,11pt]{article}
\pdfoutput=1 

\usepackage{jheppub} 

\usepackage[T1]{fontenc} 

\title{\boldmath Magnetic Impurity Inspired Abelian Higgs Vortices}

\author[a,b]{Xiaosen Han}
\author[c,d,1]{and Yisong Yang\note{Corresponding author.}}

\affiliation[a]{Institute of Contemporary Mathematics,\\School of Mathematics, Henan University,\\
Kaifeng, Henan 475004, PR China}
\affiliation[b]{Dipartimento di Matematica,\\Unversit\`{a} di Roma ``Tor Vergata",\\ Via della Ricerca Scientifica,\\ 00133 Rome, Italy (present address)}
\affiliation[c]{Department of Mathematics,\\Tandon School, New York University,\\Brooklyn, New York 11201, U. S. A}
\affiliation[d]{NYU-ECNU
Institute of Mathematical Sciences,\\New York University - Shanghai,\\3663 North Zhongshan Road, Shanghai 200062, PR China}

\emailAdd{hanxiaosen@henu.edu.cn}
\emailAdd{yisongyang@nyu.edu}

\newcommand\be{\begin{equation}}
\newcommand\ee{\end{equation}}
\newcommand\ber{\begin{eqnarray}}
\newcommand\eer{\end{eqnarray}}
\newcommand\berr{\begin{eqnarray*}}
\newcommand\eerr{\end{eqnarray*}}
\newcommand\bea{\begin{eqnarray}}
\newcommand\eea{\end{eqnarray}}

\newcommand\bfR{\mathbb{R}}\newcommand\dd{\mbox{d}}\newcommand\lm{\lambda}\newcommand\ii{\mbox{i}}
\newcommand\re{\mathrm{e}}
\newcommand\ri{\mathrm{i}}\newcommand\pa{\partial}
\newcommand{\ud}{\mathrm{d}}
\newcommand{\nn}{\nonumber}

\newcommand{\vep}{\varepsilon}\newcommand{\Om}{\Omega}

\abstract{Inspired by magnetic impurity considerations some broad classes of Abelian Higgs and Chern--Simons--Higgs  BPS vortex equations are derived and analyzed.}

\begin{document}
\maketitle
\flushbottom

\section{Introduction}

The Nielsen--Olesen vortices \cite{NO} are of important relevance in theoretical physics. In superconductivity theory,
they appear as spatially periodic topological defects, known as the Abrikosov vortices \cite{A}, allowing partial penetration of magnetic field, which characterizes the onset of  type-II superconductivity. Such vortices, when viewed in a three-dimensional formalism, give rise to string-like structure for the concentration of magnetic field, often referred to as the Nielsen--Olesen strings. These magnetic strings mediate the interaction between a monopole
and an anti-monopole, hypothetically immersed in a type-II superconductor, so that the attractive force between the
pair remains constant and the energy required to separate the pair is proportional to the distance between the
two magnetic poles. As a result, it would require infinite energy to break free a bonded pair of
a monopole and an anti-monopole placed in a type-II superconductor. This phenomenon, known as the monopole
confinement, was proposed by Mandelstam \cite{Man1,Man2}, Nambu \cite{Nambu}, and 't Hooft \cite{tH,tH1}, and has served as a profound source
of thoughts \cite{SW} for the understanding of quark confinement where the underlying Abelian Higgs model platform is extended into
various supersymmetric gauge field theory models and the classical Meissner effect is elevated into its supersymmetric
versions so that the Nielsen--Olesen magnetic strings become colored ones \cite{Auzzi,EF,EFN,EI,GJK,HT,HT2004,MY,ShY2004,ShY-vortex}. See \cite{Eto-survey,Gr,Kon-survey,Shifman-survey,ShY2,ShY,Tong} for surveys and further literature.

The full vortex equations, even in the simplest Abelian Higgs model, are too complicated to analyze to obtain a
 complete interaction picture. Fortunately, one may often extract sufficient insight
\cite{ShY,Tong} from exploring a so-called BPS structure after the works of Bogomol'nyi \cite{B} and Prasad and
Sommerfeld \cite{PS}.

In a recent study \cite{HKT}, electric and magnetic impurities are considered in the Abelian Higgs model in the context of
supersymmetric field theories.  It is shown that the usual BPS structure is preserved in the presence of such impurities.
In \cite{TW}, it is demonstrated that magnetic impurities may be viewed as heavy, frozen vortices sitting in an additional
Abelian gauge group, so that the interaction of the Abelian Higgs vortices with impurities may be described in the
framework of a product
Abelian gauge field theory containing two scalar fields $q$ and $p$ with doublet charges $(+1,-1)$ and $(0,+1)$,
respectively.

In this paper we first show that the model of Tong and Wong \cite{TW} belongs to a general product Abelian Higgs model
with two scalar fields carrying charges $(a,b)$ and $(c,d)$ for arbitrary $a,b,c,d\in\bfR$. We then show that
 this generalization may be made
to the well-known Abelian Chern--Simons--Higgs theory initially developed by Hong, Kim, and Pac \cite{HKP}, and Jackiw and Weinberg
\cite{JW}. In this latter context we  begin by extending the work \cite{HKP,JW} to include a magnetic $\sigma$-source term in the spirit of
\cite{HKT,TW}. We next expand our extension of the product Abelian Higgs model of Tong--Wong \cite{TW} to
derive a product Abelian Chern--Simons--Higgs model, under the non-degeneracy condition $ad-bc\neq0$. Our main
emphasis is to maintain the BPS structures in all these developments.

The contents of the rest of the paper is outlined as follows. In the next section, we recall some of the results in the study \cite{TW} which inspire our present work and introduce our
main notation. In Section 3, we develop our extended product Abelian gauge theory. In Section 4, we consider
the Abelian Chern--Simons--Higgs model in the presence of a magnetic source term. In Section 5, we develop
a product Abelian Chern--Simons--Higgs gauge theory. In Section 6, we present a series of existence, sometimes including uniqueness as well, results for the extended BPS vortex equations obtained. In Section 7, we draw conclusions.

\section{Abelian Higgs vortices and impurities}
\label{sec:intro}

In normalized units, the Abelian Higgs model defined over the standard Minkowski spacetime $\bfR^{2,1}$ of
signature $(+--)$, of a Bogomol'nyi--Prasad--Sommerfield (BPS) structure \cite{B,PS}, is given by the
Lagrangian density
\be\label{1.1}
{\cal L}=-\frac1{4\kappa} F_{\mu\nu}F^{\mu\nu}+D_\mu \phi \overline{D^\mu \phi}-\frac\kappa2(|\phi|^2-\zeta)^2,
\ee
where the connection or gauge-covariant derivative $D_\mu$ reads
\be
D_\mu \phi=\pa_\mu \phi-\ii A_\mu \phi,
\ee
$\phi$ is a complex-valued Higgs scalar field carrying charge $+1$, $F_{\mu\nu}=\pa_\mu A_\nu-\pa_\nu A_\mu$ the electromagnetic field induced from
the real-valued gauge field $A_\mu$ ($\mu=0,1,2$), and $\kappa,\zeta>0$ are coupling constants such that
$\sqrt{\kappa\zeta}$ gives the mass of the Higgs particle and $\sqrt{\zeta}$ measures the
energy scale of the spontaneously broken symmetry.
The equations of motion of \eqref{1.1} are
\bea
D_\mu D^\mu \phi&=&\kappa(|\phi|^2-\zeta)\phi,\\
\frac1\kappa\, \pa^\nu F_{\mu\nu}&=&\ii(\overline{\phi} D_\mu \phi-\phi\overline{D_\mu \phi}),
\eea
which are hard to solve. In the static limit, all finite-energy solutions of these equations must stay in the temporal gauge,
$A_0=0$, a statement known as the Julia--Zee theorem \cite{JZ,SY}, such that the equations can be replaced by the BPS system
\bea
F_{12}\pm\kappa(|\phi|^2-\zeta)&=&0,\\
D_1 \phi\pm\ii D_2 \phi&=&0.
\eea

In \cite{HKT,TW}, the model (\ref{1.1}) is extended to include a static magnetic source term $\sigma(x)$ ($x=(x^1,x^2)$) so that the
static energy density in the temporal gauge assumes the form
\be\label{1.6}
{\cal E}=\frac1{2\kappa} F_{12}^2+|D_j \phi |^2+\frac\kappa2(|\phi|^2-\zeta-\sigma(x))^2
\mp\sigma(x) F_{12},
\ee
whose equations of motion are
\bea
D_j D_j\phi&=&\kappa(|\phi|^2-\zeta-\sigma(x))\phi,\label{1.7}\\
\pa_j F_{12}&=&\kappa \vep^{jk}\,\ii(\overline{\phi}D_k\phi-\phi\overline{D_k\phi})\pm \kappa \pa_j\sigma.\label{1.8}
\eea
On the other hand, by the BPS quadrature procedure, we have
\bea\label{1.9}
E&=&\int_{\bfR^2}{\cal E}\,\dd x\nn\\
&=&\int_{\bfR^2}\left\{\frac1{2\kappa}(F_{12}\pm\kappa(|\phi|^2-\zeta-\sigma))^2+|D_1\phi\pm\ii D_2\phi|^2\right\}\,\dd x
\pm \zeta \int_{\bfR^2} F_{12}\,\dd x\nn\\
&\geq&2\pi\zeta |N|,
\eea
where $N$ is the winding number of the Higgs field $\phi$ which also defines the quantized flux:
\be\label{1.10}
\Phi=\int_{\bfR^2}F_{12}\,\dd x=2\pi N.
\ee
Moreover, the energy lower bound given in (\ref{1.9}) is attained when $(\phi, A_j)$ solves the BPS equations
\bea
D_1\phi\pm\ii D_2\phi&=&0,\label{1.12}\\
F_{12}\pm\kappa(|\phi|^2-\zeta-\sigma(x))&=&0,\label{1.13}
\eea
whose solutions carry the energy or vortex tension
\be
E=2\pi\zeta|N|,
\ee
which is seen to be indifferent to the magnetic source term.

It may be examined directly that (\ref{1.12})--(\ref{1.13}) imply (\ref{1.7})--(\ref{1.8}).

In \cite{TW}, Tong and Wong show how the magnetic impurities as prescribed in (\ref{1.6}) can be regarded as
heavy, frozen vortices residing in another Abelian gauge group. Such a formalism is a theory built over the
product Abelian gauge group, $\hat{U}(1)\times\tilde{U}(1)$, and accommodates two charged scalar fields, $q$ and $p$,
carrying charges $(+1,-1)$ and $(0,+1)$, respectively, so that the gauge-covariant derivatives are
\be
D_\mu q=\pa_\mu q -\ii\hat{A}_\mu q+\ii \tilde{A}_\mu q,\quad D_\mu p=\pa_\mu p-\ii\tilde{A}_\mu p,
\ee
and the Lagrangian density reads
\bea
\mathcal{L}&=&-\frac1{4\kappa}\hat{F}_{\mu\nu}\hat{F}^{\mu\nu}-\frac1{4\tilde{\kappa}}\tilde{F}_{\mu\nu}\tilde{F}^{\mu\nu}+D_\mu q\overline{D^\mu q}+D_\mu p\overline{D^\mu p}\nn\\
&&+\frac\kappa2(|q|^2-\zeta)^2+\frac{\tilde{\kappa}}2(-|q|^2+|p|^2-
\tilde{\zeta})^2,
\eea
where  $\hat{F}_{\mu\nu}=\partial_\mu \hat{A}_\nu-\partial_\nu \hat{A}_\mu$ and $\tilde {F}_{\mu\nu}=\partial_\mu \tilde{A}_\nu-\partial_\nu \tilde{A}_\mu$, and
$\tilde{\kappa}>0$ and $\tilde{\zeta}>-\zeta$ are two additional coupling constants. The vacuum state of the theory is characterized by $|q|^2=\zeta, |p|^2=\zeta+\tilde{\zeta}$, and the static equations of motion in the temporal
gauge enjoy the BPS reduction \cite{TW}:
\bea
\hat{F}_{12}\pm\kappa(|q|^2-\zeta)&=&0,\label{1.16}\\
D_1 q\pm\ii D_2 q&=&0,\\
\tilde{F}_{12}\pm\tilde{\kappa}(-|q|^2+|p|^2-\tilde{\zeta})&=&0,\\
D_1 p\pm\ii D_2 p&=&0.\label{1.19}
\eea

In the  present work our main concern is to uncover some extended classes of the BPS equations
in the same framework of Tong and Wong \cite{TW}. First we show that the above Tong--Wong BPS vortex equations belong to a broader formalism that
accommodates general doublet charges of the form $(a,b), (c,d)$ for arbitrary real parameters $a,b,c,d$.
Next we extend our study to the Abelian Chern--Simons model \cite{HKP,JW}. In this context we first derive the Chern--Simons BPS vortex equations in the presence of a magnetic impurity term in view of the studies \cite{HKT,TW}. We then
obtain BPS vortex equations for the scalar fields carrying charges $(a,b)$ and $(c,d)$ as in the Abelian Higgs model case
considered earlier, subject to the non-degeneracy condition
\be\label{1.20}
ad-bc\neq0.
\ee

\section{Generalized Abelian Higgs vortices}
\setcounter{equation}{0}

Let $\hat{A}_\mu$ and $\tilde{A}_\mu$ be real-valued gauge fields lying in the Lie algebras of $\hat{U}(1)$ and $\tilde{U}(1)$, respectively, and $\hat{F}_{\mu\nu}=\partial_\mu \hat{A}_\nu-\partial_\nu \hat{A}_\mu$ and $\tilde {F}_{\mu\nu}=\partial_\mu \tilde{A}_\nu-\partial_\nu \tilde{A}_\mu$ the induced electromagnetic fields. The
gauge-covariant derivatives on the scalar fields $q$ and $p$ carrying charges $(a,b)$ and $(c,d)$, respectively, are
defined by
\be
D_\mu q=\partial_\mu q-\ii (a\hat{A}_\mu+b\tilde{A}_\mu)q,\quad D_\mu p=\partial_\mu p-\ii (c\hat{A}_\mu+d\tilde{A}_\mu) p,\quad \mu=0,1,2,
\ee
where $a, b, c, d $ are real coupling parameters. Note that in this section we will not assume that these parameters satisfy (\ref{1.20}) for full generality.

Motivated by \cite{TW}, we consider the following Lagrangian density
\be\label{2.2}
\mathcal{L}=-\frac1{4\kappa}\hat{F}_{\mu\nu}\hat{F}^{\mu\nu}-\frac1{4\lm}\tilde{F}_{\mu\nu}\tilde{F}^{\mu\nu}+D_\mu q\overline{D^\mu q}+D_\mu p\overline{D^\mu p}-V(|q|^2, |p|^2),
\ee
where the potential density $V$  assumes  the quartic form
\be
 V(|q|^2, |p|^2)=\frac\kappa2\left(a\big[|q|^2-\xi\big]+c\big[|p|^2-\zeta\big]\right)^2+\frac{\lm}2\left(b\big[|q|^2-\xi\big]+d\big[|p|^2-\zeta\big]\right)^2,
\ee
with $\kappa,\lm>0$ being constants. The Euler--Lagrange equations of (\ref{2.2}) are
\bea
D_\mu D^\mu q&=&a\kappa (a[|q|^2-\xi]+c[|p|^2-\zeta]) q+b\lm (b[|q|^2-\xi]+d[|p|^2-\zeta])q,\\
D_\mu D^\mu p&=&c\kappa (a[|q|^2-\xi]+c[|p|^2-\zeta]) p+d\lm (b[|q|^2-\xi]+d[|p|^2-\zeta])p,\\
\frac1\kappa\pa^\nu\hat{ F}_{\mu\nu}&=&\ii(a[\overline{q} D_\mu q-q\overline{D_\mu q}]+c[\overline{p} D_\mu p-p\overline{D_\mu p}]),\\
\frac1\lm\pa^\nu \tilde{F}_{\mu\nu}&=&\ii(b[\overline{q} D_\mu q-q\overline{D_\mu q}]+d[\overline{p} D_\mu p-p\overline{D_\mu p}]),
\eea
which are complicated.

On the other hand, note that we have the identities
\bea
 |D_jq|^2&=&|D_1q\pm\ii D_2q|^2\pm\ii\left(\partial_1\big[q\overline{D_2q}\big]-\partial_2\big[q\overline{D_1q}\big]\right)\pm\big(a\hat{F}_{12}+b\tilde{F}_{12}\big)|q|^2,\label{2.4}\\
  |D_jp|^2&=&|D_1p\pm\ii D_2p|^2\pm\ii\left(\partial_1\big[p\overline{D_2p}\big]-\partial_2\big[p\overline{D_1p}\big]\right)\pm\big(c\hat{F}_{12}+d\tilde{F}_{12}\big)|p|^2.\label{2.5}
\eea
Thus, in view of (\ref{2.4}) and (\ref{2.5}), we see that in the temporal gauge, $\hat{A}_0=0,\tilde{A}_0=0$, the  static energy density  becomes
\berr
\mathcal{E}&=&\frac1{2\kappa}\hat{F}_{12}^2+ \frac1{2\lm}\tilde{F}_{12}^2+|D_jq|^2+|D_jp|^2+V(|q|^2, |q|^2)\\
&=& \frac1{2\kappa}\left(\hat{F}_{12}\pm \kappa\big(a\big[|q|^2-\xi\big]+c\big[|p|^2-\zeta\big]\big)\right)^2+\frac1{2\lm}\left(\tilde{F}_{12}\pm \lm\big(b\big[|q|^2-\xi\big]+d\big[|p|^2-\zeta\big]\big)\right)^2\\
&&\mp\hat{F}_{12}\big(a\big[|q|^2-\xi\big]+c\big[|p|^2-\zeta\big]\big)\mp\tilde{F}_{12}\big(b\big[|q|^2-\xi\big]+d\big[|p|^2-\zeta\big]\big)\\
&&+|D_1q\pm\ri D_2q|^2+|D_1p\pm\ri D_2p|^2\pm\ii\left(\partial_1\big[q\overline{D_2q}\big]-\partial_2\big[q\overline{D_1q}\big]\right)\pm\big(a\hat{F}_{12}+b\tilde{F}_{12}\big)|q|^2\\
&&\pm\ii\left(\partial_1\big[p\overline{D_2p}\big]-\partial_2\big[p\overline{D_1p}\big]\right)\pm\big(c\hat{F}_{12}+d\tilde{F}_{12}\big)|p|^2\\
&=& \frac1{2\kappa}\left(\hat{F}_{12}\pm \kappa\big(a\big[|q|^2-\xi\big]+c\big[|p|^2-\zeta\big]\big)\right)^2+\frac1{2\lm}\left(\tilde{F}_{12}\pm \lm\big(b\big[|q|^2-\xi\big]+d\big[|p|^2-\zeta\big]\big)\right)^2\\
&&+|D_1q\pm\ri D_2q|^2\pm(a\xi+c\zeta)\hat{F}_{12}\pm(b\xi+d\zeta)\tilde{F}_{12}\\
&&+|D_1p\pm\ri D_2p|^2\pm\ii\left(\partial_1\big[q\overline{D_2q}\big]-\partial_2\big[q\overline{D_1q}\big]\right)\pm\ii\left(\partial_1\big[p\overline{D_2p}\big]-\partial_2\big[p\overline{D_1p}\big]\right).
\eerr
Hence the energy has a lower bound
\berr
E=\int_{\bfR^2}\mathcal{E}\,\dd x \ge \pm(a\xi+c\zeta)\int_{\bfR^2}\hat{F}_{12}\,\dd x \pm(b\xi+d\zeta)\int_{\bfR^2}\tilde{F}_{12}\,\dd x,
\eerr
which is attained  only if the field configuration $(q,p,\hat{A}_j,\tilde{A}_j)$ satisfies the first-order equations
\ber
\hat{F}_{12}\pm \kappa\big(a\big[|q|^2-\xi\big]+c\big[|p|^2-\zeta\big]\big)=0,\label{a2}\\
\tilde{F}_{12}\pm \lm\big(b\big[|q|^2-\xi\big]+d\big[|p|^2-\zeta\big]\big)=0,\label{a3}\\
D_1q\pm \ri D_2q=0,\label{a4}\\
D_1p\pm \ri D_2p=0,\label{a5}
\eer
which are of the BPS type. The Tong--Wong equations (\ref{1.16})--(\ref{1.19}) are seen to correspond to the limiting case $a=1,b=-1,c=0,d=1$ in (\ref{a2})--(\ref{a5}). Note also that, in deriving these equations, we do not require the non-degeneracy condition (\ref{1.20}). The fulfillment of such a condition is required only when we attempt to establish an existence theory for the vortex
solutions of the equations, which will be addressed in Section 6.

\section{Chern--Simons vortices in presence of magnetic impurities}
\setcounter{equation}{0}

Adding a magnetic source term as in \cite{HKT,TW}, the Abelian Chern--Simons--Higgs Lagrangian density of \cite{HKP,JW} is modified into
\be\label{3.1}
{\cal L}=-\frac14\kappa\vep^{\mu\nu\alpha}A_\mu F_{\nu\alpha}+D_\mu \phi \overline{D^\mu\phi}-\frac1{\kappa^2}
|\phi|^2 (|\phi|^2-\sigma(x)-\zeta)^2\pm\sigma(x) F_{12},
\ee
where $\zeta>0$.
The Euler--Lagrange equations of (\ref{3.1}) are
\bea
\frac12\kappa \vep^{\mu\nu\alpha}F_{\nu\alpha}&=&\ii(\overline{\phi}D^\mu\phi-\phi\overline{D^\mu\phi}) +J^\mu_\sigma,\label{3.2}\\
D_\mu D^\mu \phi&=&-\frac1{\kappa^2}\left((|\phi|^2-\sigma-\zeta)^2+2|\phi|^2 (|\phi|^2-\sigma-\zeta)\right)\phi,\label{3.3}
\eea
where $J^\mu_\sigma$ is the $\sigma$-generated current density given as
\be
J^0_\sigma=0,\quad J_\sigma^i=\pm\vep^{ij}\pa_j\sigma.
\ee

For static solutions, the $\mu=0$ component of (\ref{3.2}) is the Chern--Simons Gauss law constraint,
\be\label{3.5}
\kappa F_{12}=2A_0|\phi|^2\equiv\rho,
\ee
which relates the induced magnetic field $F_{12}$ to the electric charge density $\rho$. Thus, in view of (\ref{3.5}), we
see that the energy density becomes
\bea
{\cal E}&=&-{\cal L}\nn\\
&=&\kappa A_0 F_{12}-A_0^2|\phi|^2+|D_j\phi|^2+\frac1{\kappa^2}|\phi|^2(|\phi|^2-\zeta-\sigma)^2\mp\sigma F_{12}\nn\\
&=&\frac{\kappa^2}{4|\phi|^2}F_{12}^2+|D_j\phi|^2+\frac1{\kappa^2}|\phi|^2(|\phi|^2-\zeta-\sigma)^2\mp\sigma F_{12}\nn\\
&=&\left(\frac\kappa2\frac{F_{12}}{|\phi|}\pm\frac{|\phi|}\kappa(|\phi|^2-\zeta-\sigma)\right)^2\mp F_{12}(|\phi|^2-\zeta)+|D_j\phi|^2.\label{3.6}
\eea
As before, we also have
\be\label{3.7}
|D_j\phi|^2=|D_1\phi\pm\ii D_2\phi|^2\pm\ii (\pa_1[\phi\overline{D_2\phi}]-\pa_2[\phi\overline{D_1\phi}])\pm F_{12}|\phi|^2.
\ee
Combining (\ref{3.6}) and (\ref{3.7}), we have
\bea\label{3.8}
E&=&\int_{\bfR^2}{\cal E}\,\dd x\nn\\
&=&\int_{\bfR^2}\left(\frac\kappa2\frac{F_{12}}{|\phi|}\pm\frac{|\phi|}\kappa(|\phi|^2-\zeta-\sigma)\right)^2\dd x
+\int_{\bfR^2}|D_1\phi\pm\ii D_2\phi|^2\,\dd x\pm \zeta\int_{\bfR^2}F_{12}\,\dd x\nn\\
&\geq&2\pi\zeta |N|,
\eea
where the winding $N$ appears again as in (\ref{1.10}) and the sign convention $|N|=\pm N$ is followed. The saturation of
the energy lower bound (\ref{3.8}) leads to the Chern--Simons BPS vortex equations
\bea
D_1\phi\pm\ii D_2\phi&=&0,\label{3.9}\\
F_{12}\pm\frac2{\kappa^2}|\phi|^2 (|\phi|^2-\zeta-\sigma(x))&=&0,\label{3.10}
\eea
which contain the classical source-free self-dual Chern--Simons equations \cite{HKP,JW} for $\sigma=0$, as anticipated.

It can be examined directly that (\ref{3.9})--(\ref{3.10}), coupled with (\ref{3.5}), imply (\ref{3.2})--(\ref{3.3}).

For an $N$-vortex solution, we have in view of (\ref{3.5}) and (\ref{3.8}) the formulas expressing the carried magnetic flux,
electric charge, and tension, as follows:
\be
\Phi=\int F_{12}\,\dd x=2\pi N,\quad Q=\int \rho\,\dd x=2\pi\kappa N,\quad E=2\pi\zeta|N|.
\ee

\section{Generalized Chern--Simons--Higgs vortices}
\setcounter{equation}{0}

We now extend our studies in the previous two sections to the Abelian Chern--Simons--Higgs situation
containing two scalar fields $q$ and $p$ of the doublet charges $(a,b)$ and $(c,d)$, respectively.
 For this purpose, consider the following Lagrangian density
\ber
\mathcal{L}=-\frac{\kappa_1}{4}\vep^{\mu\nu\alpha}\hat{A}_\mu\hat{F}_{\nu\alpha}-\frac{\kappa_2}{4}\vep^{\mu\nu\alpha}\tilde{A}_\mu\tilde{F}_{\nu\alpha}+D_\mu q\overline{D^\mu q}+D_\mu p\overline{D^\mu p}-V(|q|^2, |p|^2), \label{b1}
\eer
where the potential density $V$ is to be specified later.

The equations of motion of (\ref{b1}) read
\bea
 \frac{\kappa_1}{2}\vep^{\mu\nu\alpha}\hat{F}_{\nu\alpha}&=&\hat{j}^\mu\equiv -\ri (a[q\overline{D^\mu q}-\overline{q}D^\mu q]+c[p\overline{D^\mu p}-\overline{p}D^\mu p]),\label{b2}\\
  \frac{\kappa_2}{2}\vep^{\mu\nu\alpha}\tilde{F}_{\nu\alpha}&=&\tilde{j}^\mu\equiv -\ri (b[q\overline{D^\mu q}-\overline{q}D^\mu q]+d[p\overline{D^\mu p}-\overline{p}D^\mu p]),\label{b3}\\
  D_\mu D^\mu q&=&\frac{\partial V(|q|^2, |p|^2)}{\partial\overline{q}},\label{b4}\\
  D_\mu D^\mu p&=&\frac{\partial V(|q|^2, |p|^2)}{\partial\overline{p}}.\label{b5}
\eea

The static energy density  is given by
\bea
 \mathcal{E} &=&-{\cal L}\nn\\
&=&\kappa_1\hat{A}_0\hat{F}_{12}+\kappa_2\tilde{A}_0\tilde{F}_{12}-(a\hat{A}_0+b\tilde{A}_0)^2|q|^2-(c\hat{A}_0+d\tilde{A}_0)^2|p|^2\nn\\
&&+|D_jq|^2+|D_jp|^2+V(|q|^2, |p|^2).\label{b6}
\eea

The Gauss laws (the $\mu=0$ components of the equations \eqref{b2} and \eqref{b3}) may be read off to be
\ber
\kappa_1\hat{F}_{12}= 2a(a\hat{A}_0+b\tilde{A}_0)|q|^2+2c(c\hat{A}_0+d\tilde{A}_0)|p|^2,\label{b7}\\
\kappa_2\tilde{F}_{12}= 2b(a\hat{A}_0+b\tilde{A}_0)|q|^2+2d(c\hat{A}_0+d\tilde{A}_0)|p|^2,\label{b8}
\eer
which can be converted to yield the relations
 \ber
(a\hat{A}_0+b\tilde{A}_0)|q|^2&=&\frac{1}{2(ad-bc)}(d\kappa_1\hat{F}_{12}-c\kappa_2\tilde{F}_{12}),\label{b9}\\
(c\hat{A}_0+d\tilde{A}_0)|p|^2&=&\frac{1}{2(ad-bc)}(-b\kappa_1\hat{F}_{12}+a\kappa_2\tilde{F}_{12}),\label{b10}
 \eer
where and in the sequel we observe the non-degeneracy condition (\ref{1.20}).
Consequently, we have
\ber
&&(a\hat{A}_0+b\tilde{A}_0)^2|q|^2+(c\hat{A}_0+d\tilde{A}_0)^2|p|^2\nn\\
&&=\frac{1}{2(ad-bc)}\left\{(a\hat{A}_0+b\tilde{A}_0)(d\kappa_1\hat{F}_{12}-c\kappa_2\tilde{F}_{12})+(c\hat{A}_0+d\tilde{A}_0)(-b\kappa_1\hat{F}_{12}+a\kappa_2\tilde{F}_{12})\right\}\nn\\
&&=\frac12\left(\kappa_1\hat{A}_0\hat{F}_{12}+\kappa_2\tilde{A}_0\tilde{F}_{12}\right).\label{b11}
\eer
Thus the energy density (\ref{b6}) may be rewritten as
\ber
\mathcal{E}&=&\frac12\left(\kappa_1\hat{A}_0\hat{F}_{12}+\kappa_2\tilde{A}_0\tilde{F}_{12}\right)+|D_jq|^2+|D_jp|^2+ V(|q|^2, |q|^2)\nn\\
&=&(a\hat{A}_0+b\tilde{A}_0)^2|q|^2+(c\hat{A}_0+d\tilde{A}_0)^2|p|^2+|D_jq|^2+|D_jp|^2+ V(|q|^2, |q|^2)\nn\\
&=& \left(\frac{d\kappa_1\hat{F}_{12}-c\kappa_2\tilde{F}_{12}}{2(ad-bc)|q|}\right)^2+\left(\frac{-b\kappa_1\hat{F}_{12}+a\kappa_2\tilde{F}_{12}}{2(ad-bc)|p|}\right)^2\nn\\
&&+|D_jq|^2+|D_jp|^2+ V(|q|^2, |p|^2).\label{b12}
\eer

From (\ref{b12}) we are led to choosing the potential density
\ber
 V(|q|^2, |p|^2)&=&|q|^2\left(\left[\frac{a^2}{\kappa_1}+\frac{b^2}{\kappa_2}\right]\big[|q|^2-\xi\big]+\left[\frac{ac}{\kappa_1}+\frac{bd}{\kappa_2}\right]\big[|p|^2-\zeta\big]\right)^2\nn\\
  &&+|p|^2\left(\left[\frac{ac}{\kappa_1}+\frac{bd}{\kappa_2}\right]\big[|q|^2-\xi\big]+\left[\frac{c^2}{\kappa_1}+\frac{d^2}{\kappa_2}\right]\big[|p|^2-\zeta\big]\right)^2. \label{b13}
\eer

Therefore we obtain
\ber
\mathcal{E}&=& \left(\frac{d\kappa_1\hat{F}_{12}-c\kappa_2\tilde{F}_{12}}{2(ad-bc)|q|}\pm|q|\left(\left[\frac{a^2}{\kappa_1}+\frac{b^2}{\kappa_2}\right]\big[|q|^2-\xi\big]+\left[\frac{ac}{\kappa_1}+\frac{bd}{\kappa_2}\right]\big[|p|^2-\zeta\big]\right)\right)^2\nn\\
&&+\left(\frac{-b\kappa_1\hat{F}_{12}+a\kappa_2\tilde{F}_{12}}{2(ad-bc)|p|}\pm|p|\left(\left[\frac{ac}{\kappa_1}+\frac{bd}{\kappa_2}\right]\big[|q|^2-\xi\big]+\left[\frac{c^2}{\kappa_1}+\frac{d^2}{\kappa_2}\right]\big[|p|^2-\zeta\big]\right)\right)^2\nn\\
&&\mp\frac{1}{ad-bc}\left\{(d\kappa_1\hat{F}_{12}-c\kappa_2\tilde{F}_{12})\left(\left[\frac{a^2}{\kappa_1}+\frac{b^2}{\kappa_2}\right]\big[|q|^2-\xi\big]+\left[\frac{ac}{\kappa_1}+\frac{bd}{\kappa_2}\right]\big[|p|^2-\zeta\big]\right)\right.\nn\\
&&\left.+(-b\kappa_1\hat{F}_{12}+a\kappa_2\tilde{F}_{12})\left(\left[\frac{ac}{\kappa_1}+\frac{bd}{\kappa_2}\right]\big[|q|^2-\xi\big]+\left[\frac{c^2}{\kappa_1}+\frac{d^2}{\kappa_2}\right]\big[|p|^2-\zeta\big]\right)\right\}\nn\\
&&+|D_1q\pm\ri D_2q|^2\pm\ii\left(\partial_1\big[q\overline{D_2q}\big]-\partial_2\big[q\overline{D_1q}\big]\right)\pm\big(a\hat{F}_{12}+b\tilde{F}_{12}\big)|q|^2\nn\\
&&+|D_1p\pm\ri D_2p|^2\pm\ii\left(\partial_1\big[p\overline{D_2p}\big]-\partial_2\big[p\overline{D_1p}\big]\right)\pm\big(c\hat{F}_{12}+d\tilde{F}_{12}\big)|p|^2\nn\\
&=&\left(\frac{d\kappa_1\hat{F}_{12}-c\kappa_2\tilde{F}_{12}}{2(ad-bc)|q|}\pm|q|\left(\left[\frac{a^2}{\kappa_1}+\frac{b^2}{\kappa_2}\right]\big[|q|^2-\xi\big]+\left[\frac{ac}{\kappa_1}+\frac{bd}{\kappa_2}\right]\big[|p|^2-\zeta\big]\right)\right)^2\nn\\
&&+\left(\frac{-b\kappa_1\hat{F}_{12}+a\kappa_2\tilde{F}_{12}}{2(ad-bc)|p|}\pm|p|\left(\left[\frac{ac}{\kappa_1}+\frac{bd}{\kappa_2}\right]\big[|q|^2-\xi\big]+\left[\frac{c^2}{\kappa_1}+\frac{d^2}{\kappa_2}\right]\big[|p|^2-\zeta\big]\right)\right)^2\nn\\
&&+|D_1q\pm\ri D_2q|^2\pm\ii\left(\partial_1\big[q\overline{D_2q}\big]-\partial_2\big[q\overline{D_1q}\big]\right)\pm(a\xi+c\zeta)\hat{F}_{12}\nn\\
&&+|D_1p\pm\ri D_2p|^2\pm\ii\left(\partial_1\big[p\overline{D_2p}\big]-\partial_2\big[p\overline{D_1p}\big]\right)\pm(b\xi+d\zeta)\tilde{F}_{12}.\label{b14}
\eer
So the energy admits a lower bound
 \ber
 E=\int_{\bfR^2}\mathcal{E}\,\dd x\ge\pm(a\xi+c\zeta)\int_{\bfR^2}\hat{F}_{12}\,\dd x \pm(b\xi+d\zeta)\int_{\bfR^2}\tilde{F}_{12}\,\dd x, \label{b15}
 \eer
which is saturated only if the following first-order equations hold
 \ber
 \frac{d\kappa_1\hat{F}_{12}-c\kappa_2\tilde{F}_{12}}{2(ad-bc)}\pm|q|^2\left(\left[\frac{a^2}{\kappa_1}+\frac{b^2}{\kappa_2}\right]\big[|q|^2-\xi\big]+\left[\frac{ac}{\kappa_1}+\frac{bd}{\kappa_2}\right]\big[|p|^2-\zeta\big]\right)=0,\label{b16}\\
 \frac{-b\kappa_1\hat{F}_{12}+a\kappa_2\tilde{F}_{12}}{2(ad-bc)}\pm|p|^2\left(\left[\frac{ac}{\kappa_1}+\frac{bd}{\kappa_2}\right]\big[|q|^2-\xi\big]+\left[\frac{c^2}{\kappa_1}+\frac{d^2}{\kappa_2}\right]\big[|p|^2-\zeta\big]\right)=0, \label{b17}\\
 D_1q\pm \ii D_2q=0, \label{b18}\\
 D_1p\pm \ii D_2p=0. \label{b19}
 \eer

We note that the equations \eqref{b16}--\eqref{b17} may be rearranged such that $\hat{F}_{12}$ and $\tilde{F}_{12}$ stand out to be expressed explicitly in terms of $|q|^2$ and $|p|^2$:
 \ber
 &&\hat{F}_{12}\pm\frac{2a}{\kappa_1}|q|^2\left(\left[\frac{a^2}{\kappa_1}+\frac{b^2}{\kappa_2}\right]\big[|q|^2-\xi\big]+\left[\frac{ac}{\kappa_1}+\frac{bd}{\kappa_2}\right]\big[|p|^2-\zeta\big]\right)\nn\\
 &&\pm\frac{2c}{\kappa_1}|p|^2\left(\left[\frac{ac}{\kappa_1}+\frac{bd}{\kappa_2}\right]\big[|q|^2-\xi\big]+\left[\frac{c^2}{\kappa_1}+\frac{d^2}{\kappa_2}\right]\big[|p|^2-\zeta\big]\right)=0,\label{b20}\\
  &&\tilde{F}_{12}\pm\frac{2b}{\kappa_2}|q|^2\left(\left[\frac{a^2}{\kappa_1}+\frac{b^2}{\kappa_2}\right]\big[|q|^2-\xi\big]+\left[\frac{ac}{\kappa_1}+\frac{bd}{\kappa_2}\right]\big[|p|^2-\zeta\big]\right)\nn\\
 &&\pm\frac{2d}{\kappa_2}|p|^2\left(\left[\frac{ac}{\kappa_1}+\frac{bd}{\kappa_2}\right]\big[|q|^2-\xi\big]+\left[\frac{c^2}{\kappa_1}+\frac{d^2}{\kappa_2}\right]\big[|p|^2-\zeta\big]\right)=0.\label{b21}
 \eer
Alternatively these two equations may be rewritten usefully as
 \ber
 &&a\hat{F}_{12}+b\tilde{F}_{12}\pm2\left(\frac{a^2}{\kappa_1}+\frac{b^2}{\kappa_2}\right)|q|^2\left(\left[\frac{a^2}{\kappa_1}+\frac{b^2}{\kappa_2}\right]\big[|q|^2-\xi\big]+\left[\frac{ac}{\kappa_1}+\frac{bd}{\kappa_2}\right]\big[|p|^2-\zeta\big]\right)\nn\\
 &&\pm2\left(\frac{ac}{\kappa_1}+\frac{bd}{\kappa_2}\right)|p|^2\left(\left[\frac{ac}{\kappa_1}+\frac{bd}{\kappa_2}\right]\big[|q|^2-\xi\big]+\left[\frac{c^2}{\kappa_1}+\frac{d^2}{\kappa_2}\right]\big[|p|^2-\zeta\big]\right)=0,\label{b22}\\
  &&c\hat{F}_{12}+d\tilde{F}_{12}\pm2\left(\frac{ac}{\kappa_1}+\frac{bd}{\kappa_2}\right)|q|^2\left(\left[\frac{a^2}{\kappa_1}+\frac{b^2}{\kappa_2}\right]\big[|q|^2-\xi\big]+\left[\frac{ac}{\kappa_1}+\frac{bd}{\kappa_2}\right]\big[|p|^2-\zeta\big]\right)\nn\\
 &&\pm2\left(\frac{b^2}{\kappa_1}+\frac{d^2}{\kappa_2}\right)|p|^2\left(\left[\frac{ac}{\kappa_1}+\frac{bd}{\kappa_2}\right]\big[|q|^2-\xi\big]+\left[\frac{c^2}{\kappa_1}+\frac{d^2}{\kappa_2}\right]\big[|p|^2-\zeta\big]\right)=0.\label{b23}
 \eer

\section{Solutions of vortex equations}

In this section we present a series of rigorous existence results for the BPS vortex equations discussed in the previous
sections. In the first subsection, we consider the simplest cases, (\ref{1.12})--(\ref{1.13}) and (\ref{3.9})--(\ref{3.10}).
It is seen that, even in such a simple situation, the Abelian Higgs equations and Chern--Simons equations
exhibit some sharply different features with respect to the presence of a source term.
In the second subsection, we consider the generalized Abelian Higgs equations (\ref{a2})--(\ref{a5}). In the last
subsection, we consider the generalized Abelian Chern--Simons--Higgs equations \eqref{b16}--\eqref{b19}.

\setcounter{equation}{0}

\subsection{Vortex equations in presence of $\sigma(x)$-source terms}

We first consider the system of BPS vortex equations (\ref{1.12})--(\ref{1.13}). From (\ref{1.12}) we know that the
zeros of $\phi$ are finitely many and of algebraic multiplicities. In fact, the total number of zeros, counting
multiplicities, is the winding number of $\phi$ near infinity of $\bfR^2$ (cf. \cite{JT}). Hence, using
\be\label{6.1}
Z_\phi= \{z_1,\dots,z_N\}
\ee
to denote the set of zeros of $\phi$ (multiple zeros are counted with repetitions), the substitution $u=\ln|\phi|^2$ recasts (\ref{1.12})--(\ref{1.13}) into the
nonlinear elliptic equation
\be\label{6.2}
\Delta u=2\kappa(\re^u-\zeta-\sigma(x))+4\pi\sum_{s=1}^N \delta_{z_s}(x),
\ee
where $\delta_z$ denote the usual Dirac measure concentrated at $z$.
In \cite{JT}, an existence and uniqueness theory is developed for this equation when the $\sigma$-term is absent. The presence of
the $\sigma$-term in (\ref{6.2}), on the other hand, introduces some technical complications for obtaining an existence and uniqueness theory
under the general finite source-energy condition
\be\label{6.3}
\int_{\bfR^2}\sigma^2(x)\,\dd x<\infty.
\ee

Here, for simplicity, we consider the equation over a doubly-periodic lattice domain, $\Omega$, which may be viewed as a
flat 2-torus. That is, $\Om$ represents a lattice cell hosting periodically distributed Abrikosov vortices \cite{A}, where
periodicity is realized by the 't Hooft boundary condition \cite{tH0,WY}.

Let $u_0$ be a doubly-periodic function over $\Om$ satisfying \cite{Aubin}
\be\label{6.4}
\Delta u_0 =-\frac{4\pi N}{|\Om|}+4\pi\sum_{s=1}^N\delta_{z_s}(x),
\ee
where $|\Om|$ denotes the total area of $\Om$. Then $u=u_0+v$ transforms (\ref{6.2}) over $\Om$ into the following
source-free equation
\be\label{6.5}
\Delta v=2\kappa(\re^{u_0+v}-\zeta-\sigma(x))+\frac{4\pi N}{|\Om|},\quad x\in\Om.
\ee
Integrating (\ref{6.5}), we arrive at the natural constraint
\be
\zeta|\Om|+\int_\Om\sigma(x)\,\dd x-\frac{2\pi N}{\kappa}=\int_\Om \re^{u_0+v}\,\dd x,
\ee
which leads to the necessary condition
\be\label{6.7}
\frac{2\pi N}{\kappa}<\zeta|\Om|+\int_\Om \sigma(x)\,\dd x.
\ee
When $\sigma=0$, (\ref{6.7}) is known as the Bradlow bound \cite{Brad,MR,N}. See also \cite{N1,N2}. Using the methods in \cite{WY,ybk}, it is not hard to show that (\ref{6.7}) is also sufficient to ensure the existence of a solution, and that, the solution must be
unique. In particular,  if we decompose $\sigma$ into the sum of its positive and negative parts, $\sigma_+=\max\{\sigma,0\}$ and $\sigma_-=\max\{-\sigma,0\}$, so that
\be
\sigma=\sigma_+-\sigma_-,
\ee
then we see from the bound (\ref{6.7}) that  the number of vortices allowed is enhanced by $\sigma_+$ but diminished by $\sigma_-$, which is an interesting phenomenon.

Note that another more transparent way, perhaps, to understand (\ref{6.7}) is to rewrite it as
\be\label{6.9}
\frac{2\pi N}{\kappa|\Om|}<\zeta+\sigma_0,
\ee
where $\sigma_0$ is the average value of $\sigma(x)$ over $\Om$:
\be
\sigma_0=\frac1{|\Om|}\int_\Om\sigma(x)\,\dd x.
\ee
Thus (\ref{6.9}) spells out a necessary and sufficient condition for the average value of the source term $\sigma(x)$ in
order that an $N$-vortex solution exists over $\Om$. In particular, if $\sigma_0\leq -\zeta$, no $N$-vortex solution exists
for any $\kappa>0$ and domain $\Om$.

The existence and uniqueness of an $N$-vortex solution realizing $N$ arbitrarily prescribed vortex points given as
the zeros of the Higgs scalar $\phi$  indicates that
the set of $N$-vortex solutions of the equations  (\ref{1.12})--(\ref{1.13}) depends exactly on $2N$ parameters
labeling  those zeros.

We next consider the Chern--Simons--Higgs vortex equations (\ref{3.9})--(\ref{3.10}).

Assuming the zero set of $\phi$ is as given in (\ref{6.1}) and taking $u=\ln|\phi|^2$ as before, we reduce (\ref{3.9})--(\ref{3.10}) into
\be\label{6.11}
\Delta u=\frac4{\kappa^2}\re^u(\re^u-\zeta-\sigma(x))+4\pi\sum_{s=1}^N\delta_{z_s}(x).
\ee
If the problem is considered over a doubly-periodic domain, $\Om$, then $u=u_0+v$ (where the background function $u_0$ satisfies (\ref{6.4})) renders (\ref{6.11}) into
\be\label{6.12}
\Delta v=\frac4{\kappa^2}\re^{u_0+v}(\re^{u_0+v}-\zeta-\sigma(x))+\frac{4\pi N}{|\Om|}.
\ee
Integrating (\ref{6.12}), we arrive at the constraint
\be
4\int_\Om\left(\re^{u_0+v}-\frac{(\zeta+\sigma)}2\right)^2\,\dd x=\int_\Om (\zeta+\sigma(x))^2\,\dd x-{4\pi N\kappa^2},
\ee
similar to that in the Abelian Higgs model case, which leads us to the necessary condition
\be
\kappa^2<{\frac1{4\pi N}\int_\Om (\zeta+\sigma(x))^2\,\dd x}\equiv\eta^2_0.
\ee
Applying the methods in \cite{CY,ybk}, it can be shown that there is a critical value $0<\kappa_c<\eta_0$ such that
an $N$-vortex solution exists for any $\kappa\in(0,\kappa_c)$ but no solution exists for $\kappa>\kappa_c$. However,
when there is a solution, there is also a secondary solution \cite{T}. Thus the set of solutions of the  Chern--Simons--Higgs
$N$-vortex equations (\ref{3.9})--(\ref{3.10}) depends on at least $4N$ parameters.

It is interesting to note that, unlike in the Abelian Higgs situation, $N$-vortices exist when $\kappa$ is small enough as far as
\be
\sigma(x)\not\equiv-\zeta.
\ee

To end this subsection, we remark that a solution to (\ref{6.11}) over the full plane $\bfR^2$,
with any $\kappa>0$ and $N\geq1$, satisfying $u\to\ln\zeta$ as $|x|\to\infty$ always exists under the finite source-energy
condition (\ref{6.3}).

\subsection{Abelian Higgs vortex equations}

We now consider the BPS equations (\ref{a2})--(\ref{a5}).
The zero sets of the charged scalar fields $q$ and $p$ are denoted by
\be\label{6.16}
Z_q=\{z_1,\dots,z_{N_1}\} \quad\text{and }\quad  Z_p=\{\tilde{z}_1,\dots,\tilde{z}_{N_2}\},
\ee
respectively.
From the equations \eqref{a4}--\eqref{a5}, we have
\be\label{6.17}
a\hat{F}_{12}+b\tilde{F}_{12}=\frac12\Delta\ln|q|^2, \quad
c\hat{F}_{12}+d\tilde{F}_{12}=\frac12\Delta\ln|p|^2,
\ee  away from the zeros of $q$ and $p$.
Let $u=\ln|q|^2$ and $v=\ln|p|^2$.  Then  from  \eqref{a2}--\eqref{a3} we obtain the following coupled nonlinear elliptic
equations
\ber
 \Delta u&=& 2(\kappa a^2+\lambda b^2)\big(\re^{u}-\xi\big)+2(\kappa ac+\lambda bd)\big(\re^{v}-\zeta\big)+4\pi\sum\limits_{s=1}^{N_1}\delta_{z_s},\label{a6}\\
 \Delta v&=&2(\kappa ac+\lambda bd)\big(\re^{u}-\xi\big)+2(\kappa c^2+\lambda d^2)\big(\re^{v}-\zeta\big)+4\pi\sum\limits_{s=1}^{N_2}\delta_{\tilde{z}_s}.\label{a7}
\eer

 Let $K$ be a matrix of the form
\be K\equiv2\begin{pmatrix}
\kappa a^2+\lambda b^2&\kappa ac+\lambda bd\\
\kappa ac+\lambda bd&\kappa c^2+\lambda d^2
\end{pmatrix}. \label{a8}
\ee
Then we write the  equations \eqref{a6}--\eqref{a7}   into a compact form
\ber
\Delta\begin{pmatrix}u\\v\end{pmatrix}=K\begin{pmatrix}\re^u-\xi\\\re^v-\zeta\end{pmatrix}+4\pi\begin{pmatrix}
\sum\limits_{s=1}^{N_1}\delta_{z_s}\\ \sum\limits_{s=1}^{N_2}\delta_{\tilde{z}_s}\end{pmatrix}.\label{a9}
\eer
Since $\kappa,\lambda>0$, $ad-bc\neq0$, the matrix $K$ defined by \eqref{a8}  is positive definite. Then  we may use the a direct minimization approach developed in \cite{LY}
to establish the existence and uniqueness results for the equations \eqref{a6}--\eqref{a7} in both
planar case and doubly periodic case. These results are summarized as follows.

\begin{enumerate}
\item[(i)]   On $\mathbb{R}^2$ there is a unique solution  satisfying the boundary condition
    $
     u=\ln\xi, v=\ln\zeta
    $ at infinity.
    Moreover, this solution enjoys the exponential decay estimate
    \bea
    &&(u(x)-\ln\xi)^2+(v(x)-\ln\zeta)^2+|\nabla u(x)|^2+|\nabla v(x)|^2\nn\\
&&=
     \mbox{O}\left(\re^{-\sqrt{\lambda_0\min\{\xi, \zeta\}}|x|}\right),
      \eea
  as $|x|\to\infty$, where $\lambda_0$ is the smaller eigenvalue of the matrix
$K$
with $K$ defined in \eqref{a8}.

\item[(ii)] On a doubly periodic domain $\Omega$, there exists a solution
 if and only if
 \ber
  \frac{(\kappa c^2+\lambda d^2)N_1-(\kappa ac+\lambda bd)N_2}{\kappa\lambda(ad-bc)^2} <\frac{\xi|\Omega|}{2\pi},\label{aa10}\\
   \frac{(\kappa a^2+\lambda b^2)N_2-(\kappa ac+\lambda bd)N_1}{\kappa\lambda(ad-bc)^2} <\frac{\zeta|\Omega|}{2\pi},
     \label{a10}
 \eer
hold simultaneously. Besides, the solution is unique if it exits.

\item[(iii)]
In both cases, there hold the quantized integrals
 \ber
\int\left\{(\kappa a^2+\lambda b^2)\big(\re^{u}-\xi\big)+(\kappa ac+\lambda bd)\big(\re^{v}-\zeta\big)\right\}\ud x=-2\pi N_1,\label{6.25}\\
\int\left\{(\kappa ac+\lambda bd)\big(\re^{u}-\xi\big)+(\kappa c^2+\lambda d^2)\big(\re^{v}-\zeta\big)\right\}\ud x=-2\pi N_2,\label{6.26}
 \eer
where the integration is carried out over $\bfR^2$ or $\Om$.
\end{enumerate}

In view of  \eqref{a2}, \eqref{a3} , (\ref{6.25}), and (\ref{6.26}), we obtain the fluxes
\bea
\hat{\Phi}&=&\int\hat{F}_{12}\,\dd x=\frac{2\pi}{(ad-bc)}(d N_1 -b N_2),\label{6.28}\\
\tilde{\Phi}&=&\int\tilde{F}_{12}\,\dd x=
\frac{2\pi}{(ad-bc)}( a N_2-cN_1).\label{6.29}
\eea
Thus, using (\ref{6.28}) and (\ref{6.29}), we arrive at the exact value of the vortex tension
\be\label{E}
E=\int{\cal E}\,\dd x=2\pi (\xi N_1+\zeta N_2).
\ee

It may be checked that (\ref{aa10}) and (\ref{a10}) may be recast into the bounds

\subsection{Chern--Simons vortex equations}

Consider \eqref{b16}--\eqref{b19} so that the sets of zeros of $q$ and $p$ are as prescribed in (\ref{6.16}).
As before the equations \eqref{b18}--\eqref{b19} yield
\ber
a\hat{F}_{12}+b\tilde{F}_{12}=\frac12\Delta\ln|q|^2, \quad
c\hat{F}_{12}+d\tilde{F}_{12}=\frac12\Delta\ln|p|^2,\label{b24}
\eer  away from the zeros of $q$ and $p$.
Set  $u=\ln|q|^2$ and $v=\ln|p|^2$.  Hence by   \eqref{b22}--\eqref{b23} we obtain the following nonlinear elliptic equations
\ber
 \Delta u&=&4\left(\frac{a^2}{\kappa_1}+\frac{b^2}{\kappa_2}\right)\re^u\left(\left[\frac{a^2}{\kappa_1}+\frac{b^2}{\kappa_2}\right]\big[\re^u-\xi\big]+\left[\frac{ac}{\kappa_1}+\frac{bd}{\kappa_2}\right]\big[\re^v-\zeta\big]\right)\nn\\
 &&+4\left(\frac{ac}{\kappa_1}+\frac{bd}{\kappa_2}\right)\re^v\left(\left[\frac{ac}{\kappa_1}+\frac{bd}{\kappa_2}\right]\big[\re^u-\xi\big]+\left[\frac{c^2}{\kappa_1}+\frac{d^2}{\kappa_2}\right]\big[\re^v-\zeta\big]\right)\nn\\
&&+4\pi\sum\limits_{s=1}^{N_1}\delta_{z_s},\label{b25}\\
 \Delta v&=&4\left(\frac{ac}{\kappa_1}+\frac{bd}{\kappa_2}\right)\re^u\left(\left[\frac{a^2}{\kappa_1}+\frac{b^2}{\kappa_2}\right]\big[\re^u-\xi\big]+\left[\frac{ac}{\kappa_1}+\frac{bd}{\kappa_2}\right]\big[\re^v-\zeta\big]\right)\nn\\
 &&+4\left(\frac{b^2}{\kappa_1}+\frac{d^2}{\kappa_2}\right)\re^v\left(\left[\frac{ac}{\kappa_1}+\frac{bd}{\kappa_2}\right]\big[\re^u-\xi\big]+\left[\frac{c^2}{\kappa_1}+\frac{d^2}{\kappa_2}\right]\big[\re^v-\zeta\big]\right)\nn\\
&&+4\pi\sum\limits_{s=1}^{N_2}\delta_{\tilde{z}_s}.\label{b26}
\eer
These equations  can again be rewritten into a vector form
\ber
\Delta\begin{pmatrix}u\\v\end{pmatrix}=K\begin{pmatrix}\re^u&0\\0&\re^v\end{pmatrix}K\begin{pmatrix}\re^u-\xi\\\re^v-\zeta\end{pmatrix}+4\pi\begin{pmatrix}\sum\limits_{s=1}^{N_1}\delta_{z_s}\\ \sum\limits_{s=1}^{N_2}\delta_{\tilde{z}_s}\end{pmatrix},\label{b27}
\eer
where the matrix $K$ is defined  as
\ber K\equiv2\begin{pmatrix}
\frac{a^2}{\kappa_1}+\frac{b^2}{\kappa_2}&\frac{ac}{\kappa_1}+\frac{bd}{\kappa_2}\\[3mm]
\frac{ac}{\kappa_1}+\frac{bd}{\kappa_2}&\frac{c^2}{\kappa_1}+\frac{d^2}{\kappa_2}
\end{pmatrix} \label{b28}
\eer
Following the approach of \cite{Ycs97}, we may get an existence theorem for the   equations \eqref{b25}--\eqref{b26} over $\mathbb{R}^2$. The results are summarized as follows.

There exists a solution to the equations \eqref{b25}--\eqref{b26} satisfying the boundary condition
$
 u= \ln\xi, v= \ln\zeta$ at infinity.
 Moreover, there holds the asymptotic estimate
     \ber
     (u(x)-\ln\xi)^2+(v(x)-\ln\zeta)^2+|\nabla u(x)|^2+|\nabla v(x)|^2=
     \mbox{O}\left(\re^{-\sqrt{\lambda_0\min\{\xi, \zeta\}}|x|}\right),
      \eer
  near infinity, where $\lambda_0$ is the smaller eigenvalue of the matrix $K{\rm diag}(\xi, \zeta )K$ with $K$ being
given by (\ref{b28}).
Furthermore, for the solution obtained, there hold the quantized integrals
 \ber\label{6.36}
{\Large\int} K\begin{pmatrix}\re^u&0\\0&\re^v\end{pmatrix}K\begin{pmatrix}\re^u-\xi\\\re^v-\zeta\end{pmatrix} \ud x=-4\pi\begin{pmatrix} N_1\\N_2\end{pmatrix},
 \eer
over $\bfR^2$.

The existence problem over a doubly-periodic domain $\Om$ becomes more sophisticated in the general setting. For simplicity,
we consider the case where
 $\kappa_1=\kappa_2=\kappa$. Then, with
  \ber K_0\equiv2\begin{pmatrix}
a^2+b^2&ac+bd\\[3mm]
ac+bd&c^2+d^2
\end{pmatrix},\label{b30}
\eer
the  equations \eqref{b25}--\eqref{b26}  take the form
\ber
\Delta\begin{pmatrix}u\\v\end{pmatrix}=\frac{1}{\kappa^2}K_0\begin{pmatrix}\re^u&0\\0&\re^v\end{pmatrix}K_0\begin{pmatrix}\re^u-\xi\\\re^v-\zeta\end{pmatrix}+4\pi\begin{pmatrix}\sum\limits_{s=1}^{N_1}\delta_{z_s}\\ \sum\limits_{s=1}^{N_2}\delta_{\tilde{z}_s}\end{pmatrix},\label{b31}
\eer

Using a recently developed approach in \cite{HY} we obtain  the following existence results, assuming $ac+bd<0$ in addition, for technical reasons.

 If the  system  \eqref{b31} admits a solution   over $\Omega$, then there must hold
\ber\label{6.39}
\kappa^2(1, 1)K_0^{-1}(N_1, N_2)^\tau<\frac{|\Omega|(\xi, \zeta)K_0(\xi, \zeta)^\tau}{4\pi}.
\eer
In particular there will be no solution whatsoever when $\kappa$ is not sufficiently small.
On the other hand, there exists a suitably small positive constant $\kappa_0$ such that, when $0<\kappa<\kappa_0$, the
equation  \eqref{b31}  has a solution.
Moreover, the quantized integrals  (\ref{6.36}) are valid for any $\kappa_1,\kappa_2>0$ over $\Om$ as well.

Using (\ref{6.36}) and (\ref{b20})--(\ref{b21}), we immediately obtain the same flux formulas (\ref{6.28}) and (\ref{6.29}) as in the
extended Abelian Higgs vortex equation situation. Similarly, we see that the same vortex tension formula (\ref{E})
holds as well for the Abelian Chern--Simons vortex equations here.

\section{Conclusions}

In this paper we investigated several extended directions inspired by the recent studies
on the Abelian Higgs vortices in the presence of a magnetic source term \cite{HKT,TW} and coupled with an additional Abelian gauge
field sector \cite{TW}.

\begin{enumerate}
\item[(i)] We showed that the Abelian Higgs equations with a $\sigma$-term \cite{HKT,TW}
may be extended into the Abelian Chern--Simons--Higgs formalism \cite{HKP,JW}. In the former case,
for vortices over a periodic lattice cell, there is a
specific range for the average value of $\sigma$ which allows vortices to exist and the set of $N$-vortex solutions
depends on exactly $2N$ continuous parameters. In the latter case, however, except in the critical situation when
$\sigma=-\zeta$ where $|\phi|^2={\zeta}$ is the symmetry-breaking ground state, vortices always exist  under suitable
conditions among coupling constants, vortex number, and domain area, and the set of $N$-vortex solutions depends on at least $4N$ parameters.

\item[(ii)] We showed that the product Abelian Higgs model of Tong and Wong \cite{TW}, proposed to realize
magnetic impurities as heavy, frozen vortices residing in a secondary gauge group, described by two Higgs scalar
fields, $q$ and $p$, carrying charges $(+1,-1)$ and $(0,+1)$, respectively, may be extended into a general model in which
$q$ and $p$ carry charges $(a,b)$ and $(c,d)$ for any $a,b,c,d\in\bfR$, respectively, and that this general model also
possesses a system of BPS equations as a significant reduction from the original equations of motion.

\item[(iii)] We also showed that the formalism of Tong and Wong \cite{TW} may be extended to the context of the self-dual
or BPS Abelian Chern--Simons--Higgs theory, pioneered in the earlier works by Hong, Kim, and Pac \cite{HKP}, and Jackiw and
Weinberg \cite{JW}. Using these ideas, we developed a product Abelian Chern--Simons--Higgs model hosting two
scalar fields $q$ and $p$ carrying charges $(a,b)$ and $(c,d)$, respectively, subject to the non-degeneracy condition
$ad-bc \neq0$.

\item[(iv)] For both the extended Abelian Higgs and Chern--Simons--Higgs BPS vortex equations we have established
existence results for solutions under the non-degeneracy condition, $ad-bc\neq0$, and calculated the exact values of
the fluxes and energy or tension of a multiple vortex solution, either over a doubly-periodic domain or the full plane.

\end{enumerate}

\acknowledgments

Han was partially supported by National Natural Science Foundation of China under Grant 11201118
and the Key Foundation for Henan colleges under Grant 15A110013. Both authors were partially
supported by National Natural Science Foundation of China under Grants 11471100 and 11471099.



\end{document}